\documentclass[11pt]{article}
\newcommand{\bc}{\begin{center}}
\newcommand{\ec}{\end{center}}
\newcommand{\be}{\begin{equation}}
\newcommand{\ee}{\end{equation}}
\newcommand{\bea}{\begin{eqnarray}}
\newcommand{\eea}{\end{eqnarray}}
\newcommand{\ba}{\begin{array}}
\newcommand{\ea}{\end{array}}
\newcommand{\edc}{\end{document}}

\def\f{\varphi}

\def\g{\gamma}
\def\w{\omega}
\def\O{\Omega}

\def\t{\theta}
\def\b{\beta}
\def\G{\Gamma}

\def\e{\epsilon}
\def\s{\sigma}
\def\m{\mu}

\def\l{\lambda}
\def\L{\Lambda}
\def\k{\Upsilon}
\def\d{\partial}

\def\q{\vrule height1ex width.9ex depth.1ex}

\begin{document}
\thispagestyle{empty}

{\bf On $q$- Component Models on Cayley Tree: The General Case}
\vspace{0.3cm}

G.I. BOTIROV$^1$ and U.A. ROZIKOV$^2$

$^1${\it National University of Uzbekistan, 700174, Tashkent.\\
e-mail: botirovg@yandex.ru}

$^2${\it Institute of Mathematics, 29, F.Hodjaev str., 700143,
Tashkent, Uzbekistan. \\ e-mail: rozikovu@yandex.ru} \vspace{0.4cm}

{\bf Abstract.} In the paper we generalize  results of paper [12]
for a $q$- component models on a Cayley tree of order $k\geq 2$.
We generalize them in two directions: (1) from $k=2$ to any $k\geq
2;$ (2) from concrete examples (Potts and SOS models) of $q-$
component models to any $q$- component models (with nearest
neighbor interactions). We give a set of periodic ground states
for the model. Using the contour argument which was developed in
[12] we show existence of $q$ different Gibbs
measures for  $q$-component models on Cayley tree of order $k\geq 2$.\\[1mm]


\section{Introduction}
\large

The present paper is the continuation of our investigations (see
[12],[13]) on developing a contour method on Cayley tree. We
investigate a $q$- component spin models on Cayley tree. One of
the key problems related to such spin models is the description of
the set of Gibbs measures. The method used for the description of
Gibbs measures on Cayley tree is the method of Markov random field
theory and recurrent equations of this theory (see e.g.
[2],[8],[11]).If one considers a spin model with competing
interactions then description of Gibbs measures by the method
becomes a difficult problem.  The problem of description of Gibbs
measure has a good connection with the problem of the description
the set of ground states. Because the phase diagram of Gibbs
measures is close to the phase diagram of the ground states for
sufficiently small temperatures (see [4]-[7], [9], [10], [14] for
details). A theory of phase transitions at low temperatures in
general classical lattice (on $Z^d$) systems was developed by
Pirogov and Sinai.
 This theory is now globally known
as Pirogov-Sinai theory or contour arguments  [10], [14]-[16].

In the paper we investigate a $q$ -component models on Cayley
tree. We generalize the results of paper [12]. The organization of
the paper is  as follows. In section 2 following [12] we recall
all necessary definitions. In section 3 we generalize properties
of contours of [12] from $k=2$ to any $k\geq 2.$ In section 4 we
describe a set of ground states for the model. Section 5 devoted
to prove of existence of $q$ different Gibbs measures for any
$q$-component models with nearest neighboring interactions on
Cayley tree of order $k\geq 2$. Note that in [12] this result was
proved for Potts and SOS models on the Cayley tree of order two.

\section{Definitions}

{\bf 2.1. The Cayley tree.} The Cayley tree $\Gamma^k$ (See [1])
of order $ k\geq 1 $ is an infinite tree, i.e., a graph without
cycles, from each vertex of which exactly $ k+1 $ edges issue. Let
$\Gamma^k=(V, L, i)$ , where $V$ is the set of vertexes of $
\Gamma^k$, $L$ is the set of edges of $ \Gamma^k$ and $i$ is the
incidence function associating each edge $l\in L$ with its
endpoints $x,y\in V$. If $i(l)=\{x,y\}$, then $x$ and $y$ are
called {\it nearest neighboring vertexes}, and we write $l=<x,y>$.
 The distance $d(x,y), x,y\in V$
on the Cayley tree is defined by the formula
$$
d(x,y)=\min\{ d | \exists x=x_0,x_1,...,x_{d-1},x_d=y\in V
\ \ \mbox{such that} \ \ <x_0,x_1>,...,<x_{d-1},x_d> \}.$$

For the fixed $x^0\in V$ we set $ W_n=\{x\in V\ \ |\ \  d(x,x^0)=n\},$
$$ V_n=\{x\in V\ \ | \ \  d(x,x^0)\leq n\},\ \
L_n=\{l=<x,y>\in L \ \ |\ \  x,y\in V_n\}. \eqno (1) $$

It is known that there exists a one-to-one correspondence between
the set  $V$ of vertexes of the Cayley tree of order $k\geq 1$ and
the group $G_{k}$ of the free products of $k+1$ cyclic  groups  of
the second order with generators $a_1, a_2,..., a_{k+1}$.

{\bf 2.2. Configuration space and the model.} We consider models
where the spin takes values in the set $\Phi=\{v_1, v_2, ..., v_q\},
q\geq 2.$ A {\it configuration} $\s$ on $V$ is then defined as a
function
 $x\in V\to\s(x)\in\Phi$; the set of all configurations coincides with
$\Omega=\Phi^{V}$.

Assume on $\O$ the group of spatial shifts acts. We define  a
$F_k-${\it periodic configuration} as a configuration $\s(x)$ which
is invariant under a subgroup of shifts $F_k\subset G_k$ of finite
index. For a given periodic configuration  the index of the subgroup
is called the {\it period of the configuration}. A configuration
 that is invariant with respect to all
shifts is called {\it translational-invariant}.

The Hamiltonian  of the $q$-component  model  has the form
$$
H(\s)= \sum\limits_{<x,y>\in L}\l(\s(x),\s(y))+
 \sum\limits_{x\in V}h(\s(x))  \eqno(2)
$$
where $\l(v_i, v_j)=\l_{ij}$, $i, j = 1,...,q $ is given by  a symmetric
 matrix of
order $q\times q$ , $ h(v_j)\equiv h_j\in R $ , $ j = 1,...,q$ and
$\s\in \Omega$.

{\bf 2.3. Contours on Cayley tree}

Let $\L\subset V$ be a finite set, $\L'=V\setminus \L $ and
 $\w_\L=\{\w(x), x\in \L'\},$ $\s_\L=\{\s(x), x\in \L\}$ a given
configurations . The energy of the
configuration $\s_\L $ has the form
$$H_\L(\s_\L | \w_{\L'})=\sum_{<x,y>: x, y \in \L}\l(\s(x),\s(y))+
\sum_{<x,y>: x \in \L, y\in \L'}\l(\s(x),\w(y))+\sum_{ x \in \L}h(\s(x)).
\eqno(3) $$

Let $\w^{(i)}_{\L'}\equiv v_i$, $i=1,...,q$ be a constant configuration
outside $\L .$ For each $i$ we extend the configuration $\s_\L$ inside $\L$
to the entire tree by the $i$th constant configuration and denote
this configuration by $\s^{(i)}_\L$ and $\O^{(i)}_\L=\{\s^{(i)}_\L\}.$
Now we describe a boundary of the configuration $\s^{(i)}_\L.$

Consider $V_n$ and for a given configuration $\s^{(i)}_\L\in \O^{(i)}_\L$
denote $V^{(j)}_n\equiv V^{(j)}_n(\s^{(i)}_\L)=\{t\in V_n: \s^{(i)}_\L(t)=
v_j\}, j=1,...,q.$
Let $G^{n,j}=(V^{(j)}_n, L^{(j)}_n) $ be the graph such that
$$L^{(j)}_n=\{l=<x,y>\in L: x,y\in V^{(j)}_n\},\ \  j=1,...,q.$$
It is clear, that for a fixed $n$ the graph $G^{n,j}$ contains a finite
number
$(=m)$ of maximal  connected subgraphs $G^{n,j}_r$ i.e.
$$G^{n,j}=\{G^{n,j}_1, ..., G^{n,j}_m\}, \ \ G^{n,j}_r=(V^{(j)}_{n,r},
L^{(j)}_{n,r}), \ \ r=1,...,m.$$ Here $V^{(j)}_{n,r}$ is the set
of vertexes and $L^{(j)}_{n,r}$ the set of edges of $G^{n,j}_r.$

For a set $A$ denote by $|A|$ the number of elements in $A$.

Two edges $l_1,l_2\in L, \ \ (l_1\ne l_2)$ are called {\it nearest
neighboring edges} if $|i(l_1)\cap i(l_2)|=1$, and we write $<l_1,
l_2>_1.$

For any connected component $K\subset \G^k$ denote by $E(K)$ the set
of edges of $K$ and
$$b(K)=\{l\in L\setminus E(K): \exists l_1\in E(K) \ \ \mbox{such that}
\ \ <l,l_1>_1\}. $$

DEFINITION 1. An edge $l=<x,y>\in L_{n+1}$ is called a {\it boundary
edge} of the configuration $\s^{(i)}_{V_n}$ if
 $\s^{(i)}_{V_n}(x)\ne \s^{(i)}_{V_n}(y) .$
The set of boundary edges of the configuration is called {\it boundary}
$\d (\s^{(i)}_{V_n})\equiv \G$ of this configuration.

The boundary $\G$ consists of ${q(q-1)\over 2}$ parts
$$\d_\e(\s^{(i)}_{V_n}) \equiv \G_\e, \ \ \e\in \{ij: i<j ; i ,j=1,...,q\}
\equiv Q_q,
$$
where, for instance $\G_{12}$ is the set of edges $l=<x,y>$ with
$\s(x)=v_1$ and $\s(y)=v_2.$

The (finite) sets $b(G^{n,j}_r), j=1,...,q, r=1,...,m $ (together
with indication for each edge of this set which part
$\G_\e, \e\in Q_q$ of the boundary contains this edge) are called
{\it subcontours} of the boundary $\G .$

The set $V^{(j)}_{n,r}, \ \ j=1,...,q, r=1,...,m$ is called the {\it
interior} Int$b(G^{n,j}_r)$ of  $b(G^{n,j}_r)$.

The set of edges from a subcontour $\g$ is denoted by supp$\g$ .
The configuration $\s^{(i)}_{V_n}$ takes the same value $v_j, j=1,...,q$
at all points of the connected component $G^{n,j}_r$. This value $v=
v(G^{n,j}_r)$ is called the {\it mark} of the subcontour and denoted by
$v(\g)$, where $\g=b(G^{n,j}_r).$

The collection of subcontours $\tau=\tau(\s^{(i)}_{V_n})=\{\g_r\}$
generated by the boundary $\G=\G(\s^{(i)}_{V_n})$ of the configuration
$\s^{(i)}_{V_n}$ has the following properties

(a) Every subcontour $\g\in \tau$ lies inside the set $V_{n+1}.$

(b) For every two subcontours $\g_1, \g_2\in \tau$ their supports supp$\g_1$
and supp$\g_2$ satisfy $|$supp$\g_1 \cap $supp$\g_2 |\in \{0, 1\}.$

The subcontours $\g_1, \g_2\in \tau$ are called {\it adjacent} if
$|$supp$\g_1 \cap $supp$\g_2 |= 1.$

(c) For any two adjacent subcontours $\g_1, \g_2\in \tau$ we have
$v(\g_1)\ne v(\g_2).$

A set of subcontours $A\subset \tau$ is called {\it connected } if
for any two subcontours $\g_1, \g_2\in A$ there is a sequence of
subcontours $\g_1={\tilde \g} _1, {\tilde \g} _2,..., {\tilde \g}
_l=\g_2$ in the set $A$ such that for each $i=1,...,l-1$ the
subcontours ${\tilde \g}_i$ and ${\tilde \g}_{i+1}$ are adjacent.

DEFINITION 2. Any maximal connected set (component) of subcontours
is called {\it contour} of boundary $\G .$

Let $\k=\{\g_r, r=1,2,...\}$ (where $\g_r$ is subcontour) be a contour
of $\G$ denote
$$ \mbox{Int}\k=\cup_j\mbox{Int}\g_j ;\ \
\mbox{supp}\k=\cup_j\mbox{supp}\g_j ;\ \  |\k|=|\mbox{supp}\k|.$$

\section{Properties of the contours}

For $A\subset V$ denote $\d(A)=\{x\in V\setminus A: \exists y\in A,
\ \ \mbox{such that} \ \ <x,y>\}.$

Let $G$ be a graph, denote the vertex and edge set of the graph $G$
by $V(G)$ and $E(G),$  respectively.

 LEMMA 3. {\it Let $K$ be a connected subgraph of the Cayley tree $\G^k$
such that $|V(K)|=n$, then $|\d(V(K))|=(k-1)n+2.$}

{\it Proof.} We shall use the induction over $n .$ For $n=1$ and 2
the assertion is trivial. Assume for $n=m$ the lemma is true i.e
from $|K|=m$ follows $|\d (K)|=(k-1)m+2.$ We shall prove the
assertion for $n=m+1$ i.e. for $\tilde{K}=K\cup \{x\}.$  Since
$\tilde{K}$ is connected graph we have $x\in \d (K)$ and there is
unique $y\in S_1(x)=\{u\in V: d(x,u)=1\}$ such that $y\in K.$ Thus
$\d (\tilde{K})=(\d( K)\setminus \{x\})\cup (S_1(x)\setminus
\{y\}).$ Consequently,
$$|\d (\tilde{K})|=|\d (K)|-1+k=(k-1)(m+1)+2.$$ \hfill$\q $

 LEMMA 4. [3] {\it Let $G$ be a countable graph of maximal degree
$k+1$ (i.e. each $x\in V(G)$ has at most $k+1$ neighbors)
and let $\tilde{N}_{n,G}(x)$ be the number of connected subgraphs $
G'\subset G$ with $x\in V(G')$ and $|E(G')|=n$. Then
$$\tilde{N}_{n,G}(x)\leq (e\cdot k)^n.$$}

For $x\in V$ we will write $x\in \k$ if there is $l\in \k$
such that $x\in i(l).$

Denote  $N_r(x)=|\{\k: x\in \k, |\k|=r \}|$ .

 LEMMA 5. {\it For any $k\geq 2$ we have
 $$ N_r(x)\leq \t \cdot \alpha^r, \eqno (4)$$
where $\alpha=(2ke)^{k\over k-1}$, $\t={1\over
2\sqrt[k]{\alpha}(\alpha-1)}$.}

{\it Proof.}  Denote by $K_\k$ the minimal connected subgraph of
$\G^k$, which contains a contour $\k$. It is easy to see that if
$\k=\{\g_1,...,\g_m\} , m\geq 1$, then
$$E(K_\k)=\mbox{supp}\k\cup \bigg(\cup_{i=1}^m\{<x,y>: x, y \in
\mbox{Int}\g_i\}\bigg). \eqno(5)$$

Using the fact that if $K$ is a connected subgraph of $\G^k$ then
the number of edges of $K$ equal to $|K|-1$, equality
$\sum^m_{i=1}|\g_i|=|\k|+m-1$ and  Lemma 3 we get
$$|E(K_\k)|=|\k|+ \sum_{i=1}^m (|\mbox{Int}\g_i|-1)=
|\k|+ \sum_{i=1}^m ({|\g_i|-2\over k-1}-1)={k\over k-1}
|\k|-{km+1\over k-1}. \eqno(6)$$

Since $\k\subseteq K_\k$ we get $|\k| \leq |E(K_\k)|={k\over k-1}
|\k|-{km+1\over k-1}.$ Consequently, $1\leq m \leq {|\k|-1\over k}.$
A combinatorial calculations show that
$$N_r(x)\leq \sum_{m=1}^{[{r-1\over k}]}
{|K_\k|-1 \choose r} \tilde{N}_{|K_\k|-1, \G^k}(x), \eqno(7)$$ where
$[a]$ is the integer part of $a$. Using inequality ${n \choose r}
\leq 2^{n-1}, \ \ r\leq n$ and lemma 4 from (7) we get (4).
\hfill$\q$

\section{Ground states}

For $l=<x,y>\in L$ and configuration $\s\in \O$ denote
$\s_l=\{\s(x), \s(y)\}.$  Define the energy of the configuration
$\s_l=\{v_i, v_j\}$ by

$$U(\s_l)\equiv U_{ij}\equiv \l_{ij}+{1\over k+1}(h_i+h_j).\eqno(8)$$

Then our Hamiltonian can be written as
$$H(\s)=\sum_{l\in L}U(\s_l).$$

  For a pair of configurations $\s$ and $\f$ that
coincide almost everywhere, i.e. everywhere except for a finite
number of positions, we consider a relative Hamiltonian $H(\s,\f)$,
the difference between the energies of the configurations $\s, \f$
of the form

$$
H(\s,\f)=H(\s)-H(\f)=\sum_{l\in L}(U(\s_l)-U(\f_l)).
$$

DEFINITION 6. A periodic configuration $\f$ is called {\it ground
state} (for the relative hamiltonian $H$) if  $H(\f,\s)\leq 0$ for
any configuration $\s$ that coincides with $\f$ almost everywhere.

 LEMMA 7. {\it For any normal subgroup $F_k$
with index $r$, $r\leq q$ of $G_k$ there exist at least ${q!\over
(q-r)!}$ of $F_k-$ periodic configurations.}

{\it Proof.} Since $F_k$ is the subgroup of index $r$ in $G_k,$ the
quotient group has the form $G_k/F_k=\{F_{k,0},...,F_{k,r-1}\}$ with
the coset $F_{k,0}=F_k.$ A $F_k-$periodic configuration $\s_{F_k}$
can be defined as $\s_{F_k}(x)=v_i$ if $x\in F_{k,i},$
$i=0,...,r-1.$ We have at least ${q \choose r}\cdot r!={q!\over
(q-r)!}$ possibility to define such configuration combining the
values $v_1,...,v_q.$ This completes the proof.\hfill$\q $

{\it Remark.} If $r>q$ then one can construct a $F_k-$ periodic
configuration. But in this case one has to set values of the
configuration the same on some cosets.

 The following very simple lemma gives periodic ground states.

LEMMA 8. {\it A $F_k-$ periodic configuration $\s$ is a ground state
if $U(\s_l)=U^{\rm min}$ for any $l=<x,y>\in L,$ where} $ U^{\rm
min}=\min\{U_\e: \e\in Q_q\}.$

\section{ Non uniqueness of Gibbs measure}

In this section we assume
$$U_{ii}=U^{\min}<U_\e,\ \ i=1,...,q; \e\in Q_q \eqno (9)$$ thus
the ground states of the model will be  all constant configurations
$\s^{(m)}=\{\s^{(m)}(x)=v_m, x\in V\},$ $m=1,...,q.$ Now we shall
prove that every such ground state generates a Gibbs measure.

 The energy $H_{\L}(\s |\f)$ of the configuration $\s$
in the presence of boundary configuration $\f=\{\f(x), x\in V\setminus \L\}$
is expressed by the formula
$$H_{\L}(\s |\f)=\sum_{l=<x,y>: x,y\in \L}U(\s_l)
+\sum_{l=<x,y>: x\in \L, y\in V\setminus\L}U(\s_l). \eqno (10)$$

Following lemma gives a contour representation of the Hamiltonian

LEMMA 9. \textit{The energy $H_n(\s_n)\equiv H_{V_n}(\s_n|
\f_{V'_n}=v_i)$  has the form
$$H_n(\s_n)=\sum_{\e\in Q_q}(U_\e-U_{ii})|\G_\e|+(|V_{n+1}|-1)U_{ii}, \eqno(11)$$
where $|\G_\e|$ is defined in the subsection 2.3.}

{\it Proof.} We have
$$H_n(\s_n)=\sum_{l\in
L_{n+1}}U(\s_{n,l})=\sum_{\e\in
Q_q}U_\e|\G_\e|+(|V_{n+1}|-1-|\G|)U_{ii}.\eqno(12)$$

Now using equality $|\G|=\sum_{\e\in Q_q}|\G_\e|$ from (12) we get
(11). \hfill$\q $

The Gibbs measure on the space $\O_{\L}=\{v_1,...,v_q\}^{\L}$ with
boundary condition $\f$ is defined as
$$\m_{\L,\b}(\s/\f)\equiv \m^{\f}_{\L,\b}(\s)={\bf Z}^{-1}(\L,\b,\f)
\exp(-\b H_{\L}(\s |\f)), \eqno(13)$$ where ${\bf Z}(\L, \b, \f)$ is
the normalizing factor (statistical sum).

Denote ${\mathbf U}=\{U_\e: \e\in Q_q\}$, $U^{\min}=\min_{\e\in
Q_q}U_\e$ and
$$\l_0=\min\bigg\{{\mathbf U}\setminus \{U_\e: U_\e=U^{\min}\}\bigg\}-
U^{\min}.\eqno(14)$$

LEMMA 10. \textit{Assume that (9) satisfied. Let $\g$ be a fixed
contour and
$$p_i(\g)={\sum_{\s_n:\g\subset  \G}\exp\{-\b H_n(\s_n)\}\over
\sum_{{\tilde \s}_n}\exp\{-\b H_n({\tilde \s}_n)\}}.$$
 Then
$$p_i(\g)\leq \exp\{-\b\l_0|\g|\}, \eqno(15)$$
where $\l_0$ is defined by formula (14) and $\b={1\over T}, T>0-$
temperature.}

{\it Proof.}
 Put $\O_\g=\{\s_n: \g\subset \G\}$, $\O_\g^0=\{\s_n: \g\cap
\G=\emptyset\}$  and define a map $\chi_\g:\O_\g \to \O_\g^0$ by

$$\chi_\g(\s_n)(x)=\left\{\begin{array}{ll}
v_i & \textrm{if \ \ $ x\in {\rm Int}\g$}\\
\s_n(x)& \textrm{if \ \ $x\notin {\rm Int}\g$} \\
\end{array}\right.
$$

For a given $\g$ the map $\chi_\g$ is one-to-one map.  For any
$\s_n\in \O_{V_n} $  we have
$$|\G_\e(\s_n)|=|\G_\e(\chi_\g(\s_n))|+|\g_\e|, \e\in Q_q, \eqno(16)$$
here $\g_\e=\g\cap \G_\e.$

Using Lemma  9 we  have
$$p_i(\g)={\sum_{\s_n\in \O_\g}\exp\{-\b \sum_{\e\in Q_q}(U_\e-U_{ii})|\G_\e(\s_n)|\}\over
\sum_{{\tilde \s}_n}\exp\{-\b \sum_{\e\in
Q_q}(U_\e-U_{ii})|\G_\e({\tilde \s}_n)|\}}\leq$$
$${\sum_{\s_n\in \O_\g}\exp\{-\b
\sum_{\e\in Q_q}(U_\e-U_{ii})|\G_\e(\s_n)|\}\over \sum_{{\tilde
\s}_n\in \O_\g}\exp\{-\b \sum_{\e\in
Q_q}(U_\e-U_{ii})|\G_\e(\chi_\g({\tilde \s}_n))|\}}.\eqno(17)$$

By the assumption (9) we have  $U_\e-U_{ii}\geq \l_0$ for any $\e\in
Q_p,  i=1,...,q.$ Thus using this fact and (16) from (17) we get
(15). \hfill$\q $

Using Lemmas 5 and 10 by very similar argument of [12] one can prove

{\bf Theorem 11.} \textit{If (9) satisfied  then for all
sufficiently large $\b$ there are at least $q$ Gibbs measures for
the model (2) on Cayley tree of order $k\geq 2.$}

 \vskip 0.3 truecm

 {\bf Acknowledgments.} The work supported by NATO
Reintegration Grant : FEL. RIG. 980771. The final part of this work
was done within the Junior Associate scheme  at the ICTP, Trieste,
and UAR thanks ICTP for providing financial support and all
facilities. The work also partically supported by Grants $\Phi M.
1.152$ and $\Phi.2.1.56$ of CST of the Republic Uzbekistan.

\vskip 0.3 truecm
{\bf References}

1. Baxter, R. J.: {\it Exactly  Solved Models in Statistical Mechanics},
Academic Press, London/New York, 1982.

2. Bleher, P. M. and  Ganikhodjaev, N. N.: On pure phases of the
Ising model on the Bethe lattice, {\it Theor. Probab. Appl.} {\bf
35} (1990), 216-227.

3. Borgs, C.: {\it Statistical physics expansion methods in
combinatorics and computer science},  http://research.microsoft.com/
$\sim$borgs/CBMS.pdf, March 22,  2004

4. Fern$\acute{\textrm {a}}$ndez R.: {\it Contour ensembles and the
description of Gibbsian probability distributions at low
temperature.}
 www.univ-rouen.fr/LMRS/persopage/Fernandez, 1998.

5.  Holsztynski, W. and  Slawny, J.: Peierls condition and the
number of ground states, {\it Commun. Math. Phys.} {\bf 61} (1978),
177-190 .

6.  Kashapov, I. A.: Structure of ground states in three-dimensional
Ising model with tree-step interaction, {\it Theor. Math. Phys.}
{\bf 33} (1977), 912-918 .

7.  Minlos, R. A.: {\it Introduction to mathematical statistical
physics} University lecture series, ISSN 1047-3998; v.19,  2000

8.  Mukhamedov, F. M. and Rozikov, U. A.: On Gibbs measures of
models with competing ternary and binary interactions and
corresponding von Neumann algebras. I,II. {\it Jour. of Stat.Phys.}
{\bf 114} (2004), 825-848; {\bf 119} (2005), 427-446.

9.  Peierls, R.: On Ising model of ferro magnetism. {\it Proc.
Cambridge Phil. Soc.} {\bf 32}: (1936), 477-481 .

10. Pirogov, S. A. and Sinai, Ya. G.: Phase diagrams of classical
lattice systems,I. {\it Theor. Math. Phys.} {\bf 25} (1975),
1185-1192 ; {\bf 26} (1976), 39-49.

11.  Rozikov, U. A. and Suhov, Yu. M.: A hard-core model on a Cayley
tree: an example of a loss network, {\it Queueing Syst.} {\bf 46}
(2004), 197-212.

12.  Rozikov, U. A.: On $q-$ component models on Cayley tree:
contour method, {\it Letters in Math. Phys.} {\bf 71} (2005), 27-38.

13.  Rozikov, U. A.: A  constructive description of ground states
and Gibbs measures for Ising model with two-step interactions on
Cayley tree, {\it Jour. of Stat. Phys.} {\bf 122} (2006), 217-235.

14. Sinai, Ya. G.: {\it Theory of phase transitions: Rigorous
Results} Pergamon, Oxford, 1982.

15. Zahradnik, M.: An alternate version of Pirogov-Sinai theory.
{\it Comm. Math. Phys.} {\bf 93} (1984), 559-581.

16. Zahradnik, M.: A short course on the Pirogov-Sinai theory, {\it
Rendiconti Math. Serie VII} {\bf 18} (1998), 411-486.

\end{document}